# Starlink Generation 2 Mini Satellites: Photometric Characterization

Anthony Mallama, Richard E. Cole, Scott Harrington,
Andreas Hornig, Jay Respler, Aaron Worley and Ron Lee

2023 June 11

Correspondence: anthony.mallama@gmail.com

**Abstract**

Starlink Generation 2 Mini satellites are fainter than Gen 1 spacecraft despite their larger size. The mean of apparent magnitudes for satellites in brightness mitigation mode is 7.06 +/- 0.10. When these magnitudes are adjusted to a uniform distance of 1,000 km that mean is 7.87 +/- 0.09. The brightness mitigation mode reduces distance-adjusted satellite luminosity by a factor of 12 relative to spacecraft that are not mitigated.

**1. Introduction**

The current model of Starlink Generation 2 satellites has a surface area more than four times as great as Gen 1 satellites. They are called Minis because regular Gen 2 spacecraft will be even larger. SpaceX has received regulatory approval to orbit 7500 second generation satellites. The luminosity and the number of Starlink spacecraft is a concern for researchers making celestial observations (Mroz et al., 2022). They are also a potential distraction for amateur astronomers and others wishing to view the starry sky in its natural state (Mallama and Young, 2021). The International Astronomical Union recently established a [Centre](#) to address satellite interference.

SpaceX has responded to the concerns of astronomers by developing an improved reflective layer for their Gen 2 satellites which directs more of the sunlight away from observers on the ground. The company also [explains](#) that they adjust the attitude of the solar arrays to minimize brightness when near the Earth's terminator. Their goal is to "make its second-generation satellites invisible to the naked eye when they are on station".

This paper reports on brightness measurements of Mini satellites obtained at visible wavelengths through 2023 June 3. Section 2 describes the spacecraft and their orbits. Section 3 addresses the methods of photometric measurement. Section 4 characterizes satellite luminosity. Section 5 discusses our findings in the context of related studies and Section 6 lists the conclusions.

**2. Satellite dimensions and orbits**

The area of the Mini spacecraft antenna panel is 11.07 $m^2$ while the two solar arrays add another 104.96 $m^2$. The total of 116.03 $m^2$ is 4 times the size of Gen 1 Starlinks and even exceeds that of the highly luminous BlueWalker 3 satellite (Mallama et al. 2023).

The first 21 of these spacecraft from Starlink launch 6-1 were placed into orbit on 2023 February 27 at an initial altitude of 370 km. Some of them remained at about that height or descended while others began orbit raising.

On March 22 the CEO of SpaceX [tweeted](#) "Lot of new technology in Starlink V2, so we're experiencing some issues, as expected. Some sats will be deorbited, others will be tested thoroughly before raising altitude above Space Station". That statement explains the variety of height changes.

Three of the spacecraft from launch 6-1 ascended to 480 km by late April and have remained at that altitude. SpaceX informed us on May 16 that those satellites were in their on-station conops for brightness mitigation. So, they became a priority for the observing program.

The satellites from launch 6-2 on 2023 April 19 initially orbited at 350 km. Some of those satellites descended but 8 of them reached 480 km in early May as shown in Figure 1. More satellites raised their orbits to the on-station altitude on later dates while some of those already at 480 km ascended to 520 km. We also prioritized observation of the on-station satellites from this launch.

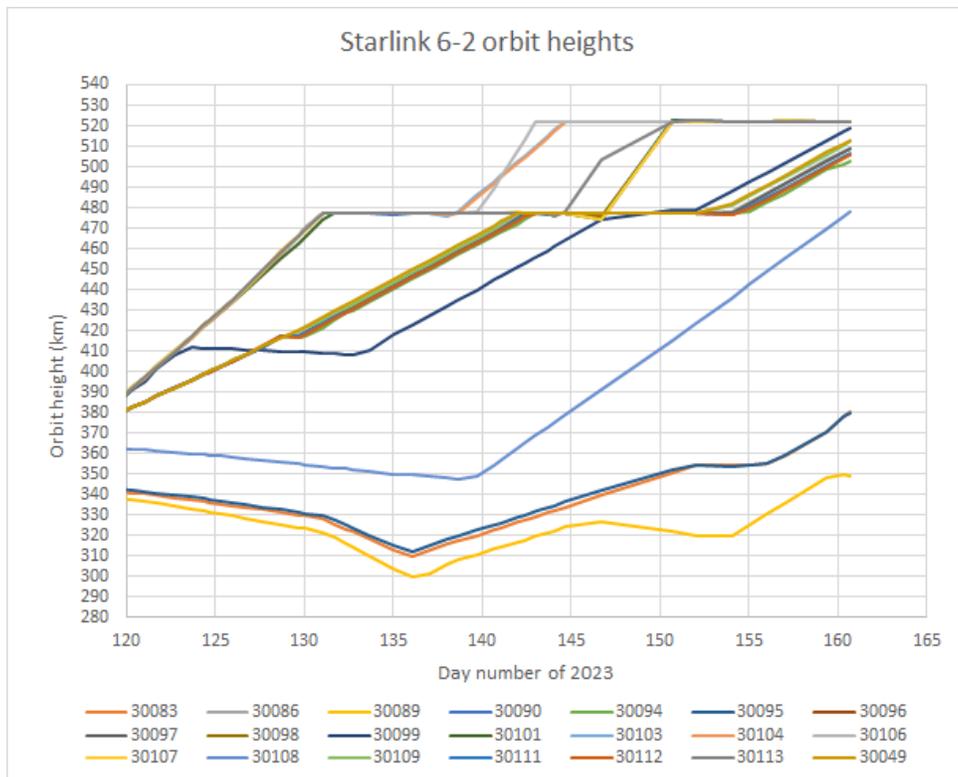

*Figure 1. The heights of satellites from Starlink launch 6-2 plotted versus time.*

Launch 6-3 occurred on May 19 with the satellites initially at 320 km. While 6-1 and 6-2 consisted of 21 spacecraft each, this launch included 22. All except one of the satellites are currently ascending.

## 3. Observational methods and data processing

The visible band observations analyzed in this study were obtained perceptually by eye or were recorded with a video camera. Satellite magnitudes are determined by comparing their brightness to nearby reference stars. Angular proximity between the spacecraft and those stars takes account of variations in sky transparency and sky brightness. The perceptual method of observing is described more thoroughly by Mallama (2022). The visual observers listed in Table 1 recorded 397 magnitudes for this study.

Table 1. Observers, methods and coordinates

```
Cole         Visual    50.552N     4.735W    100m
Harrington   Visual    36.062N    91.688W    185m
Hornig       Camera    41.878N    22.501E    400m
Lee          Visual    38.948N   104.561W   2082m
Mallama      Visual    38.982N    76.763W     42m
Respler      Visual    40.330N    74.445W    170m
Worley       Visual    41.474N    81.519W    351m
```

Video observations were obtained with a Sony Alpha A7s-I camera and a Sony FE 1.8/50 lens. The Astrometry.net application was run on a Raspberry Pi 4 device for extracting information about the stars. Specially coded python software was executed on a Windows computer to perform the overall measurements and data processing.

The magnitudes derived from video observations are equivalent to the electronic V-band of the Johnson-Cousins photometric system and to visual magnitudes obtained by eye as illustrated by Figure 2. Video frames were averaged over 5 second time intervals to generate 109 magnitudes for this study. These observations were recorded by Hornig at the coordinates listed in Table 1. More details about the video hardware and software are available [online](online).

Data processing consists of adjusting the apparent (that is, observed) magnitudes to a uniform distance and computing their phase angles. The apparent magnitudes are transformed to a distance of 1,000 km according to the inverse square law of light. The phase angle is that arc measured at the satellite between the directions to the Sun and to the observer. The next section describes how magnitudes and phase angles are used to study satellite brightness.

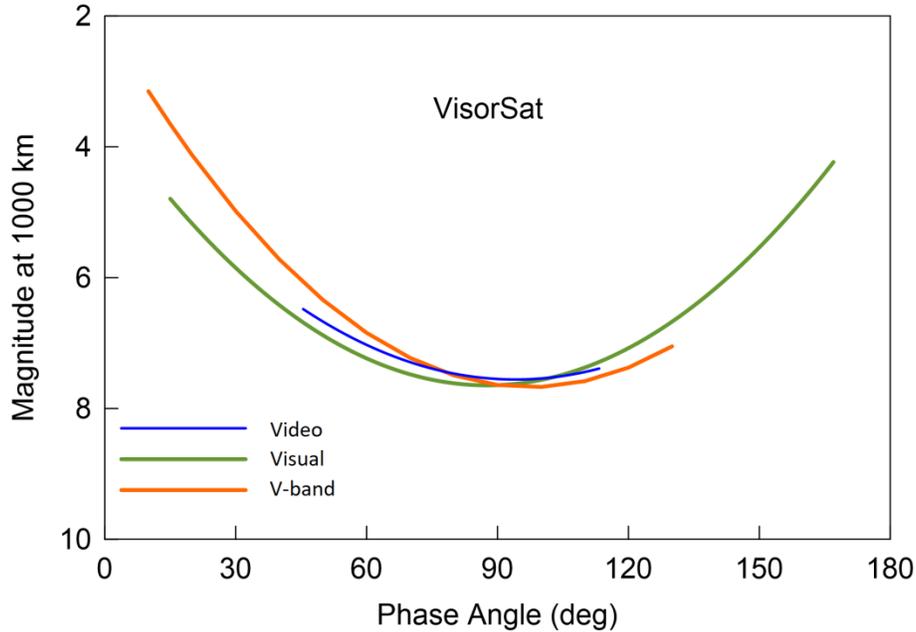

*Figure 2. The video magnitudes reported in this paper are equivalent to visual and V-band values. The lines represent least-squares fits to Starlink VisorsSat brightness as a function of phase angle. The visual phase function is from Mallama and Respler (2022) and that of the V-band is from Mallama (2021).*

## 4. Brightness characterization

Apparent magnitude is a direct measure of a satellite's interference with astronomical observations. Magnitudes brighter than 7 are generally considered problematic for research (see the discussion in Tyson et al. 2020). The distribution of apparent brightness for all Mini satellite observations is shown in Figure 3. The peak is at magnitude 2.0 but the function has a long tail stretching to magnitude 9 and its mean value is 4.46 +/- 0.10.

When the apparent magnitudes are adjusted to a common distance of 1,000 km the distribution peaks at magnitude 3.5, spans from 2.5 through 9.5, and has a mean of 5.52 +/- 0.08. The apparent and adjusted values indicate that Starlink Gen 2 Mini satellites can be very bright and are a concern for astronomers.

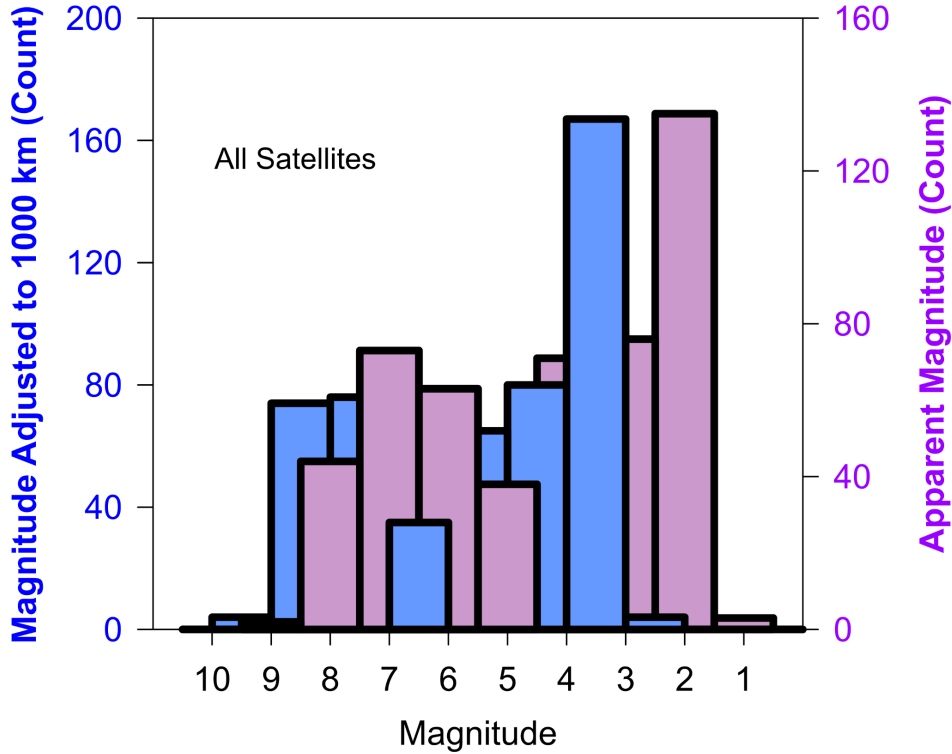

*Figure 3. The distribution of brightness for all Mini satellite observations. The peak of apparent brightness is at magnitude 2.0 and the corresponding peak for distance-adjusted values is 3.5.*

This high luminosity was expected because many of the satellite observations were obtained during early mission phases including orbit raising. SpaceX had indicated that the satellites could be visible to the unaided eye during that time. The mean of apparent magnitudes recorded exclusively during early mission is 3.97 +/- 0.09 and the corresponding mean of magnitudes adjusted to the uniform distance is 5.08 +/- 0.08. These values are brighter than the means listed in the previous paragraphs which were derived from observations of all spacecraft.

Meanwhile, the satellites that had ascended to 480 km were in their on-station brightness mitigation mode as mentioned in Section 2. Thus, the spacecraft body and solar arrays were oriented to reduce brightness. Figure 4 demonstrates that the distributions of magnitudes for the on-station spacecraft are much fainter than those described above.

The peak of the apparent magnitudes for satellites in the brightness mitigating mode is at 6.5, making them difficult or impossible to see with the unaided eye, and the peak for the distance-adjusted magnitudes is 8.0. The means are 7.06 +/- 0.10 for apparent magnitudes and 7.87 +/- 0.09 for distance-adjusted magnitudes. The adjusted mean indicates a factor of 12 reduction in luminosity relative to the corresponding mean for satellites not in brightness mitigation mode.

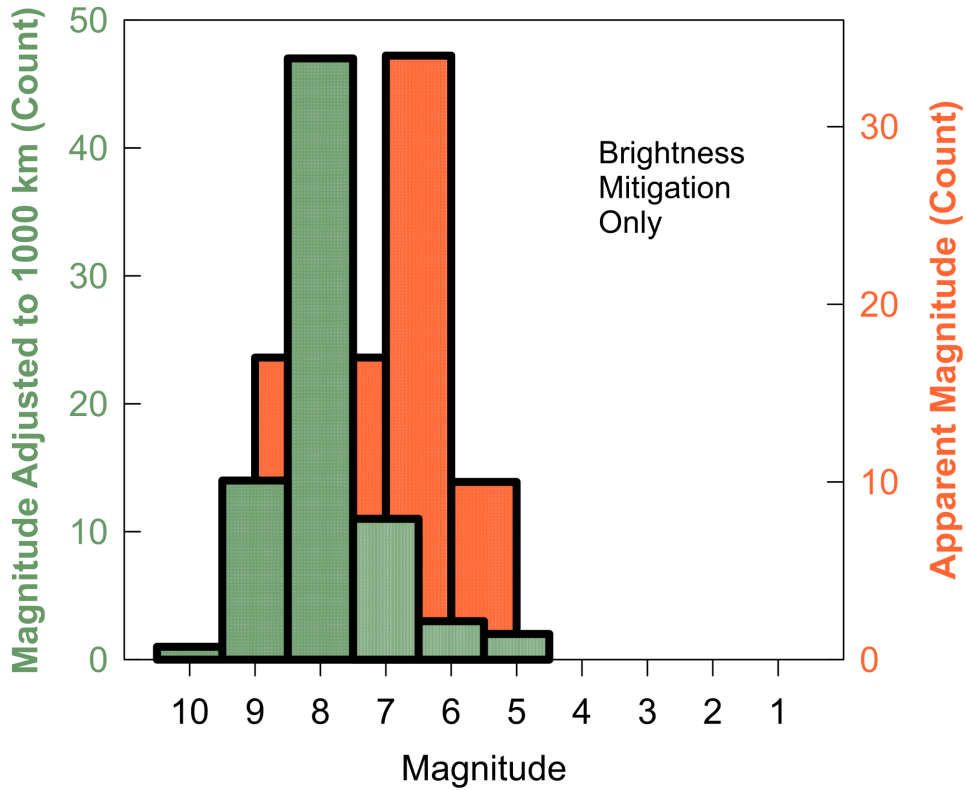

*Figure 4. The distribution of apparent brightness for on-station Mini satellites peaks at magnitude 6.5. The peak for distance-adjusted brightness is 8.0. Compare this to Figure 3.*

Table 2. Magnitude means, standard deviations and standard deviations of the means

|  | -- Unmitigated -- | | | --- Mitigated --- | | | ------ All ------ | | |
| --- | --- | --- | --- | --- | --- | --- | --- | --- | --- |
|  | **Mean** | **StDev** | **SD_Mean** | **Mean** | **StDev** | **SD_Mean** | **Mean** | **StDev** | **SD_Mean** |
| **Apparent** | 3.97 | 1.96 | 0.09 | 7.06 | 0.91 | 0.10 | 4.46 | 2.15 | 0.10 |
| **1000-km** | 5.08 | 1.70 | 0.08 | 7.87 | 0.79 | 0.09 | 5.52 | 1.89 | 0.09 |

A final method for characterizing brightness is the phase function which represents distance-adjusted brightness versus phase angle. Figure 5 illustrates this function for Mini satellites in brightness mitigation mode. Spacecraft observed at small angles are front-lit by the Sun and appear brighter than those seen at mid-range angles around 90° which are side-lit. Meanwhile, the function beyond 90° suggests greater brightness at large angles which indicates significant forward scattering of sunlight from back-lit satellites. Overall, the phase function resembles that of VisorSat as seen in Figure 2.

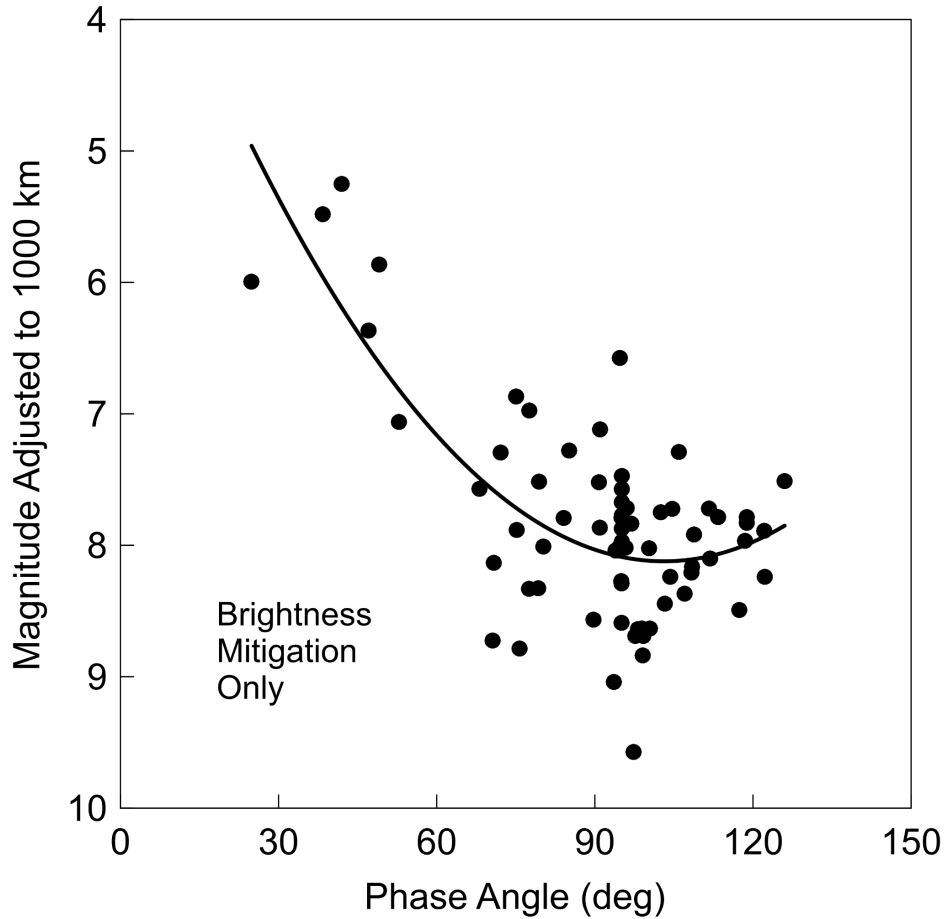

*Figure 5. The phase function for Mini satellites where a polynomial is fitted to magnitudes by least-squares. The quadratic coefficients for orders 0, 1 and 2 are: 2.630, 0.1065 and -0.0005167.*

## 5. Discussion

This study supplements earlier photometric research on the first generation of Starlink satellites. The visual brightness for all three types of Gen 1 spacecraft has been characterized by Mallama and Respler (2022). Other large-scale studies of those satellites have been published by Krantz et al. (2022) and Mallama (2021). The findings in those papers are brought up to date here with results for Gen 2 Mini spacecraft.

We find that Gen 2 satellites are very luminous during their early mission phases before reaching on-station altitudes and being placed in a brightness mitigation mode. This was expected due to their large size and because SpaceX stated before launch that "V2 Mini satellites may be somewhat bright initially, especially during orbit raising and initial operations".

On-station satellites are found to be much fainter as shown in Figure 6. A private communication from SpaceX explained that Gen 2 Starlinks place the solar arrays edge-on to the Earth's limb so observers on the ground will not usually see portions of the solar array that are illuminated by the Sun. The factor of 12 reduction in brightness that we determined for on-station Mini satellites exceeds the factor of 10 dimming found by Mallama and Respler (2023) for Gen 1 satellites in knife-edge configuration.

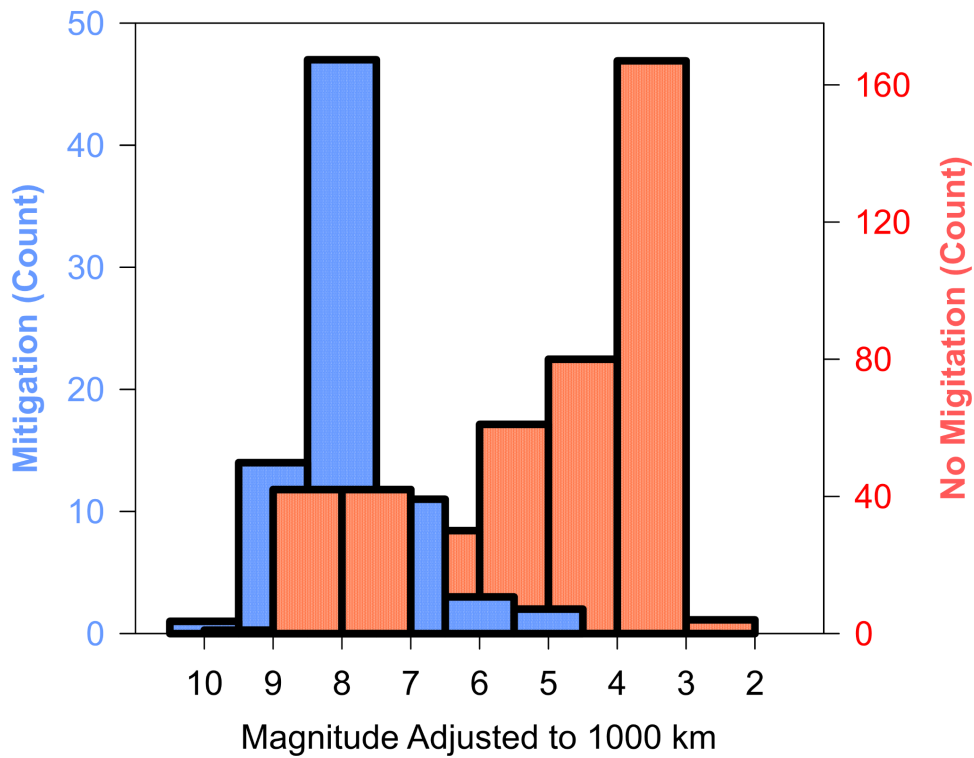

*Figure 6. The distribution of magnitudes for satellites with and without brightness mitigation applied.*

The brightness of Mini satellites may be compared to that obtained from observations of Gen 1 Starlink spacecraft (Mallama and Respsler, 2022) and to the very luminous BlueWalker 3 (BW3) (Mallama et al. 2023). Table 3 lists means of their distance-adjusted magnitudes at 1,000 km. The relative luminosities and the luminosities per unit of satellite surface area are normalized to those of BW3 which is the highest in both categories. The Mini spacecraft are the faintest and they have the least brightness per surface area by a wide margin.

Table 3. Brightness comparison

```
                               Mag_1000_km         Area      Luminosity
                              Mean   StD_Mean*      m²       Rel.   Per_Area
Starlink Gen 2 Mini           7.87     0.09       116.0      0.012    0.007
Starlink Gen 1 VisorSat       7.00     0.04        26.3      0.027    0.067
Starlink Gen 1 Post-VisorSat  6.34     0.05        26.3      0.050    0.123
Starlink Gen 1 Original       6.08     0.04        26.3      0.064    0.156
BlueWalker 3**                3.09     0.04        64.4      1.000    1.000

* StD_Mean is the standard deviation of the mean
** The values for BW3 are derived from 'regular' magnitudes and do not
include 'faint period' data. The satellite's zenith side was sometimes
tipped toward the Sun to increase insolation on its solar array and
that tilt temporarily reduced the satellite's brightness.
```

For the future, more observations will be recorded when the Mini satellites have risen to their final on-station altitude of 560 km. We also plan to study the brightness of these spacecraft with a bidirectional reflectance function (BRDF). Cole (2020, 2021) modeled the luminosity of Starlink VisorSat with a BRDF. Meanwhile, Fankhauser et al. (2023) developed a model which includes Earthshine and applied it to Starlink version 1.5.

The remainder of this discussion section addresses unusual luminosities and uncertain magnitudes. First, the brightness of some satellites surged briefly during observation. These 'flare' events occurred less than 1% of the time and they do not significantly affect our conclusions.

Second, satellites were too faint to be seen during 18 of the 506 attempted observations. Omitting these data would make average magnitudes too bright. So, 0.5 was added to the limiting magnitude during each of these observations to provide an estimate of the satellite luminosity. The limit corresponds to the faintest visible star which was generally about magnitude 8. In 6 cases where the limiting magnitude was much brighter than 8 the observation was omitted from further analysis because the satellite magnitude could not be reliably approximated.

Some observations of on-station satellites were made at ranges exceeding 1,000 km. These are grouped with brightness mitigated magnitudes. However, it is not known at what maximum distance from the Earth's terminator solar panels are oriented for mitigation.

Finally, there were two instances where Starlink-30051 on-station was as luminous as an unmitigated spacecraft. These observations are grouped with those for satellites not in the brightness reducing mode. The high luminosity may be explained by the following [statement](#) from SpaceX: "The satellites periodically need to perform burns on station to maintain their position in orbit and avoid collisions, and they will be brighter during such operations."

## 6. Conclusions

This study reports photometry of Starlink Generation 2 Mini satellites. Magnitudes were recorded during early orbit and on-station phases. Luminosity is characterized and the effectiveness of brightness mitigation is evaluated.

The mean of apparent magnitudes for Mini satellites recorded during early mission phases is 3.07 +/- 0.09 and the corresponding mean of magnitudes adjusted to a uniform distance of 1,000 km is 5.08 +/- 0.08. The means for satellites in on-station operational mode for brightness mitigation are apparent magnitude 7.06 +/- 0.10 and 7.87 +/- 0.09 for distance-adjusted magnitudes. The distance-adjusted means indicate that brightness mitigation efficiency is a factor of 12. Therefore, the Mini satellites are fainter than Gen 1 Starlink satellites despite their larger sizes.

**Acknowledgements**

The [Heavens-Above](#) web site was used to predict satellite passes. The [SeeSat](#) message board contained informative postings about satellite brightness.

**References**

Cole, R.E. 2020. A sky brightness model for the Starlink 'Visorsat' spacecraft – I, Research Notes of the American Astronomical Society, 4, 10, https://iopscience.iop.org/article/10.3847/2515-5172/abc0e9

Cole, R.E. 2021. A sky brightness model for the Starlink 'Visorsat' spacecraft. https://arxiv.org/abs/2107.06026

Fankhauser, F., Tyson, J.A. and Askari, J. 2023. Satellite optical brightness. https://arxiv.org/abs/2305.11123.

Krantz, H., Pearce, E.C. and Block, A. 2022. Characterization of LEO satellites with all-sky photometric signatures. https://arxiv.org/abs/2210.03215.

Mallama, 2021. Starlink satellite brightness – characterized from 100,000 visible light magnitudes. https://arxiv.org/abs/2111.09735.

Mallama, 2022. The method of visual satellite photometry. https://arxiv.org/abs/2208.07834.

Mallama, A. and Respler, J. 2022. Visual brightness characteristics of Starlink Generation 1 satellites. https://arxiv.org/abs/2210.17268.

Mallama, A. and Respler, J. 2023. Roll angle adjustment dims Starlink satellites. https://arxiv.org/abs/2303.01431.

Mallama, A., Cole, R.E. Tilley, S., Bassa, C. and Harrington, S. 2023. BlueWalker 3 satellite brightness characterized and modeled. https://arxiv.org/abs/2305.00831

Mallama, A. and Young, M. 2021. The satellite saga continues. Sky and Telescope, vol. 141, June, p. 16.

Mroz, P., Otarola, A., Prince, T.A., Dekany, R., Duev, D.A., Graham, M.J., Groom, S.L., Masci, F.J. and Medford, M.S. 2022. Impact of the SpaceX Starlink satellites on the Zwicky Transient Facility survey observations. https://arxiv.org/abs/2201.05343.

Tyson, J.A., Ivezić, Ž., Bradshaw, A., Rawls, M.L., Xin, B., Yoachim, P., Parejko, J., Greene, J., Sholl, M., Abbott, T.M.C., and Polin, D. (2020). Mitigation of LEO satellite brightness and trail effects on the Rubin Observatory LSST. arXiv e-prints, arXiv:2006.12417.